\documentclass[preprint,12pt]{elsarticle}

\usepackage{amsmath}
\usepackage{amssymb} 
\usepackage{cases}
\usepackage{url}
\usepackage{color}

\newcommand{\eq}[1]{Eq.(\ref{#1})}

\newcommand{\fig}[1]{Fig.\ref{#1}}

\newcommand{\Fig}[1]{Figure \ref{#1}}

\newcommand{\scl}[2]{s_{#1}^{(#2)}}
\newcommand{\wvl}[2]{w_{#1}^{(#2)}}

\journal{Optics Communications}

\begin{document}

\begin{frontmatter}



\title{Fast, large-scale hologram calculation in wavelet domain}


\author[a]{Tomoyoshi Shimobaba\corref{cor1}}
\cortext[cor1]{Tel: +81 43 290 3361; fax: +81 43 290 3361}
\ead{shimobaba@faculty.chiba-u.jp}
\author[b]{Kyoji Matsushima}
\author[a]{Takayuki Takahashi}
\author[a]{Yuki Nagahama}
\author[a]{Satoki Hasegawa}
\author[a]{Marie Sano}
\author[a]{Ryuji Hirayama}
\author[a]{Takashi Kakue}
\author[a]{Tomoyoshi Ito}

\address[a]{Chiba University, Graduate School of Engineering, 1--33 Yayoi--cho, Inage--ku, Chiba, Japan, 263--8522}
\address[b]{Department of Electrical and Electronic Engineering, Kansai University, 3--3--35 Yamate--cho, Suita, Osaka 564--8680, Japan}

\begin{abstract}
We propose a large-scale hologram calculation using WAvelet ShrinkAge-Based superpositIon (WASABI), a wavelet transform-based algorithm.
An image-type hologram calculated using the WASABI method is printed on a glass substrate with the resolution of $65,536 \times 65,536$ pixels and a pixel pitch of $1 \mu$m.
The hologram calculation time amounts to approximately 354 s on a commercial CPU, which is approximately 30 times faster than conventional methods.
\end{abstract}


\end{frontmatter}

Holograms can be calculated by simulating a light wave emitted from three-dimensional (3D) objects. 
The development of an ideal 3D display can be expected when these holograms are displayed on a spatial light modulator (SLM) \cite{poon2006digital}.
Unfortunately, this 3D display has the disadvantages that the size and viewing area of the 3D objects are small and narrow.
This has been a hindrance to their practical use; however, in recent years, several techniques to address these issues have been proposed. 
A common factor in these techniques is that to observe the 3D objects with a wide viewing angle and large size by multiple observers, it is necessary to calculate a hologram with a large number of pixels and display it spatially \cite{active,sasaki, poly} or temporally \cite{takaki}. 
In other techniques, hologram printers capable of printing large-scale holograms with minute pixel pitch have been actively employed to reproduce 3D objects with a wide viewing area and large size \cite{printer1,printer2,printer3,printer4,printer}.

The hologram calculation time for reconstructing 3D objects with wide viewing angles and large size is considerably lengthy.
Various fast hologram calculation methods have been proposed with procedures of 3D object expression. 
In the case of 3D objects represented by polygons, tilted diffraction calculation \cite{poly1} can effectively compute a hologram.
A holographic stereogram \cite{ichihashi, hs, rs,zhang2015fully} can be used when 3D objects are expressed in multi-view images. 
In the case that 3D objects are represented by an RGB-D image, a hologram calculation using diffraction calculation of cross sectional images generated from the RGB-D images has been proposed \cite{dep1} . 

For 3D objects represented by point light sources, a hologram can be calculated by adding light waves from each point light source on the hologram plane. 
Many acceleration methods for the point light source model have been proposed, suich as an image hologram method \cite{image_hol}, look-up table methods (LUT) \cite{lut1,lut2}, and wavefront recording plane methods \cite{wrp,tsang2011holographic}. 
Recently, we proposed WAvelet ShrinkAge-Based superpositIon (WASABI), a wavelet transform-based method \cite{wasabi,wasabi2}. 
This method superimposes light waves of point light sources in a wavelet domain. 
However, although the WASABI method has shown that hologram calculation can be efficiently performed using a small-scale hologram of $2,048 \times 2,048$ pixels, thus far, the implementation of large-scale hologram calculations has not been accomplished.

In this study, we generate a large-scale hologram calculation using the WASABI method. 
An image-type hologram calculated using the WASABI method is printed on a glass substrate. 
The resolution of the hologram is $65,536 \times 65,536$ pixels with the pixel pitch of $1 \mu$m. 
The calculation time of the hologram is approximately 354 s on a commercial CPU.
The calculation is capable of a 30-factor acceleration compared to conventional methods.

\section{Large-scale hologram calculation using WASABI method}
A hologram can be calculated by adding light waves emitted from each object point on the hologram plane. 
The following equation is used for the calculation:
\begin{equation}
u(x_h,y_h)=\sum_j^N a_j \exp(i \frac{2 \pi}{\lambda} r_{hj}) =\sum _j^N a_j u _{z_j}(x_h-x_j, y_h-y_j),
\label{eqn:cgh}
\end{equation}
where $i=\sqrt{-1}$, $N$ is the total number of the point light sources, $(x_h,y_h)$ is the coordinate of the hologram, $(x_j,y_j,z_j)$ and $a_j$ are the coordinate and the light intensity of $j$-th point light source, respectively.
$\lambda$ is the wave length, $r_{hj}$ is the distance between $j$-th object point and $(x_h,y_h)$, and $u_{z_j}$ is the point spread function (PSF) at the position $z_j$ of an object point.
Although this calculation is simple, when the PSF is distributed over the entire hologram, the calculation complexity is $O(NN_h^2)$ where the number of pixels of the hologram is $ N_h \times N_h $. 
Therefore, this accounts for the greatest calculating cost among hologram calculations.

\begin{figure}[htbp]
\centering
\fbox{\includegraphics[width=13cm]{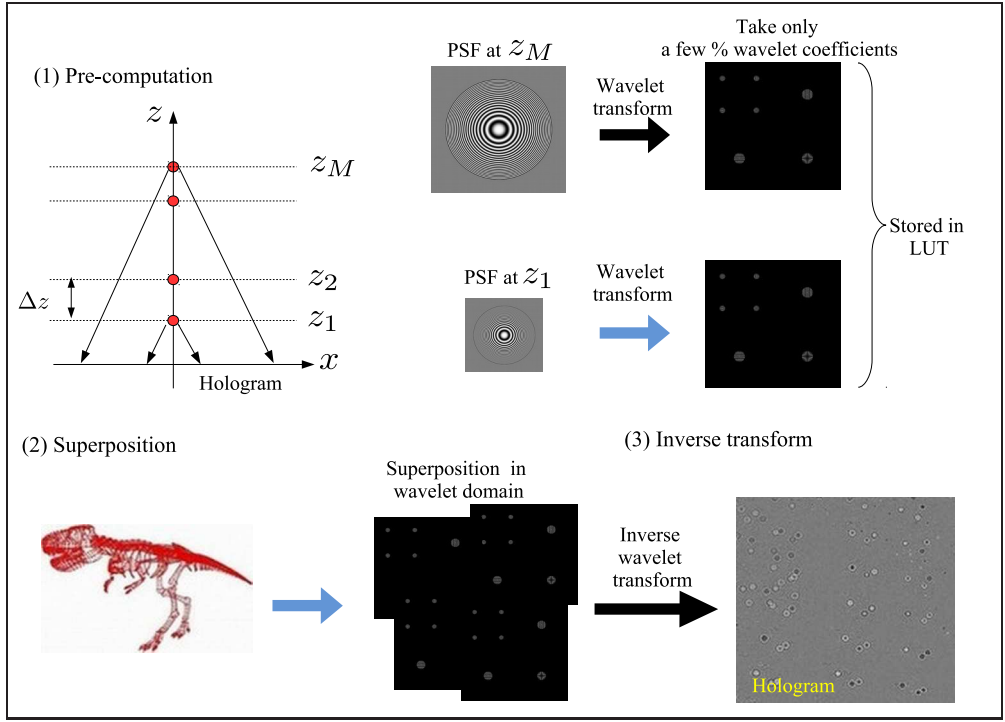}}
\caption{Calculation procedure of the WASABI method.}
\label{fig:wasabi_calculation}
\end{figure}

The WASABI method can reduce this computation amount via applying the wavelet transform to the PSF and by superposing the PSFs in the wavelet domain \cite{wasabi}. 
\Fig{fig:wasabi_calculation} shows the calculation steps of the WASABI method.
The calculation steps are explained as follows:
\begin{enumerate}
\item Pre-computation: We pre-compute PSFs by $u_{z_j}=\exp(ikr_{hj}) {\rm circ}(\sqrt{x_h^2+y_h^2}/(2W_j))$ where $W_j$ is the PSF radius of the $j$-th object point and ${\rm circ}( c ) = 1$ when $c<1$, otherwise 0. 
The depth direction of the pre-calculated PSFs are discretized by $\Delta z=z_{j+1}-z_j$ and the PSFs are converted to wavelet-domain PSFs using fast wavelet transform (FWT). 
We select $N_r$ strong coefficients among the wavelet-domain PSF and store them in the LUT. In this study, $N_r$ is determined by $N_r = \pi (W_j / 2)^2 \gamma$ where $\gamma$ is the selectivity.
\item Superposition: The superposition corresponding to \eq{eqn:cgh} in the wavelet domain is performed using the coefficients stored in the LUT.
\item Inverse transformation: The superposed result is converted to the space domain using the inverse FWT.
\end{enumerate}

\begin{figure}[htbp]
\centering
\fbox{\includegraphics[width=13cm]{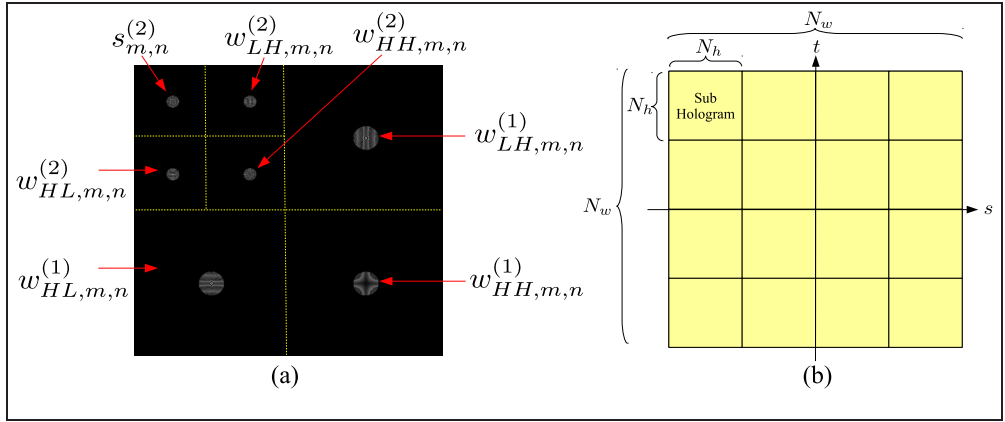}}
\caption{Expression of PSF in wavelet domain and large-scale hologram calculation: (a) PSF in wavelet domain and (b) large-scale hologram calculation.}
\label{fig:wasabi}
\end{figure}

As shown in \fig{fig:wasabi}(a), a wavelet-domain PSF is expressed by the scaling coefficient $\scl{m,n}{\ell}$ and three wavelet coefficients $\wvl{LH, m,n}{\ell}$, $\wvl{HL, m,n}{\ell}$ and $\wvl{HH, m,n}{\ell}$, where $\ell$ denotes the level of FWT and the subscripts $L$ and $H$ denote low and high frequencies, respectively.
\Fig{fig:wasabi}(a) shows an example of a wavelet-domain PSF at level 2.
In this study, we used coiflets as the wavelets at level 3.

The superposition in the wavelet domain is performed by
\begin{equation}
\psi(m,n) = \sum_{j=0}^{N} a_j \sum_{k=0}^{N_r-1} c_{z_j, k} \delta(m-\alpha_{z_j,k} x_j, n-\alpha_{z_j,k} y_j),
\label{eqn:super}
\end{equation}
where $(m,n)$ is the coordinate in the wavelet domain, $\delta(m,n)$ is the Dirac delta function and $c_{z_j, k} \in \{ \scl{m,n}{\ell}, \wvl{LH, m,n}{\ell}, \wvl{HL, m,n}{\ell}, \wvl{HH, m,n}{\ell} \}$ is the $N_r$ strong coefficients and $\alpha_{z_j,k}$ is the shift weight according to the level $\ell$ of the FWT. 

When calculating a large-scale hologram by the WASABI method, we first divide an entire hologram into sub-holograms, as shown in \fig{fig:wasabi}(b).
This division can considerably reduce the memory usage of a computer as compared with calculating the entire hologram at once.
Herein, the number of pixels of the entire hologram is $N_w \times N_w$, and the number of pixels for each sub hologram is $N_h \times N_h$. 
The index of the sub holograms currently being calculated is represented by $(s, t)$. The calculation performed is explained as follows:
\begin{enumerate}
\item Pre-computation: As described above, PSFs for a sub hologram with $N_h \times N_h$ pixels are converted to the wavelet domain via FWT. $N_r$ strong coefficients are stored in the LUT. 
\item Coordinate transformation of object points: The coefficients after FWT are defined for sub-holograms ($N_h \times N_h$ pixels), and not for the entire hologram ($N_w \times N_w$ pixels). Therefore, the positions of all the object points are shifted relative to the current sub-hologram $(s, t)$. Thus, the shift is given as $x'_j = x_j - s N_h$ and $y'_j = y_j - t N_h$.
\item Superposition: Using $x'_j$ and $y'_j$ instead of $x_j$ and $y_j$ in \eq{eqn:super}, the superposition is performed by $\psi(m,n) = \sum_{j=0}^{N} a_j \sum_{k=0}^{N_r-1} c_{z_j, k} \delta(m-\alpha_{z_j,k} x'_j, n-\alpha_{z_j,k} y'_j)$. 
\item Inverse transformation: We obtain the sub-hologram at $(s,t)$ via inverse FWT of $\psi(m,n)$. We move to the next sub-hologram.
\end{enumerate}
The large-scale hologram calculation is executed by performing the above steps 2 to 4 for all sub-holograms.

\section{Result}
We show the performance of a large hologram calculation with $N_w \times N_w=65,536 \times 65,536$ pixels of a 3D object with approximately 100,000 object points (95,949 points) using a conventional method and the WASABI method. 
We used the N-LUT method \cite{lut2} as the conventional method. 

\begin{figure}[htbp]
\centering
\fbox{\includegraphics[width=13cm]{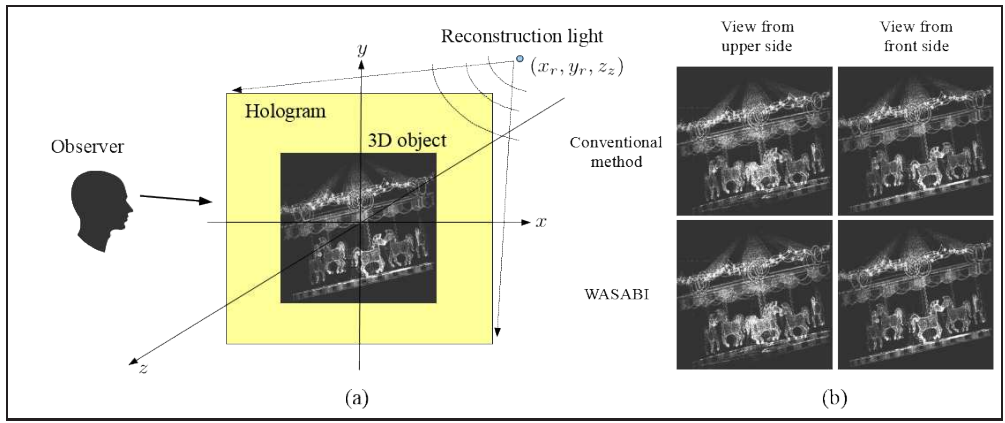}}
\caption{Setup of hologram calculation and numerical reconstructions from the hologram: (a) setup of hologram calculation and (b) numerical reconstructions.}
\label{fig:setup_num}
\end{figure}

\Fig{fig:setup_num}(a) shows the setup of the hologram calculation.
The hologram is recorded as an image-type hologram, and the 3D object is placed at +15mm to -15mm before and after the hologram plane.
The depth of the 3D object is divided by $N_z=29$, such that the depth interval $\Delta z$ is $\Delta z=z_{j+1}-z_j=30{\rm mm} / 29 \approx 1.03{\rm mm}$.
The sub-holograms have $N_h \times N_h=8,192 \times 8,192$ pixels.
The reference light is a spherical light placed at $(x_r, y_r, z_r)$=(0mm, 30mm, -400mm).
The sampling pitch of the hologram is 1$\mu$m.
The calculated hologram has a complex amplitude. In this study, we convert the complex amplitude hologram to a binary amplitude hologram because we use a hologram printer of Ref.\cite{printer}.

\Fig{fig:setup_num}(b) shows numerical reconstructions from the hologram calculated using the conventional and the WASABI methods.
The left figure of \fig{fig:setup_num}(b) shows the reconstructed images when looking at the hologram from the upper side, and the right figure is reconstructed images when looking at the hologram from the front side.
Motion parallax can be confirmed from these reconstructions. 

\begin{figure}[htbp]
\centering
\fbox{\includegraphics[width=13cm]{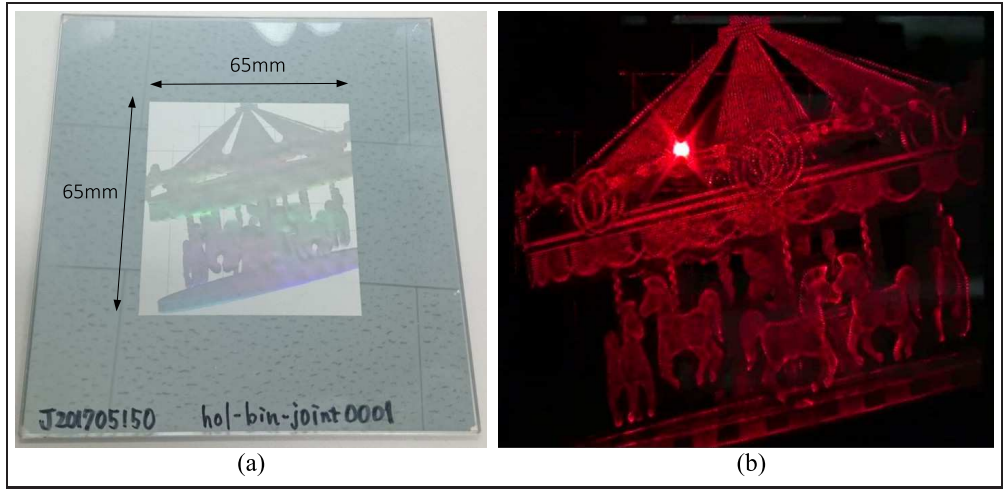}}
\caption{Hologram printed on a glass substrate and an optical reconstruction of the hologram: (a) printed hologram and (b) optical reconstruction (\textcolor{blue}{see Visualization 1}).}
\label{fig:hol_reconst}
\end{figure}

\Fig{fig:hol_reconst}(a) shows a hologram calculated via the WASABI method, which is printed on a glass substrate using the hologram printer \cite{printer}. The hologram area is 65mm $\times$ 65mm. 
\Fig{fig:hol_reconst}(b) shows the optical reconstruction of the printed hologram.
We used a LED light with a wavelength of 632.8nm as the reconstruction light.
The video in supplementary material is recorded while moving a video camera left and right. 
The viewing angle is approximately 20 degrees.

We compare the calculation time of the hologram using the conventional method \cite{lut2} and the WASABI method.
We used an Intel Xeon E5 v3 CPU with 128-GB memory and the Microsoft Visual C++ 2017 as the C++ compiler. 
In the conventional method, the calculation time is approximately 10,533 s, whereas the WASABI method is capable of calculating the hologram in approximately 354 s. 
Thus, the WASABI method is approximately 30 times faster than the conventional method.
All of the calculations were performed using our wave optics library, CWO++ \cite{shimobaba2012computational}.

\section{Conclusion}
We executed a large-scale hologram calculation using the WASABI method and compared the calculation time between the conventional and the WASABI methods.
An image-type hologram with $65,536 \times 65,536$ pixels was calculated using the WASABI method, and we confirmed the optical 3D reconstruction from the hologram printed on a glass substrate. 
The WASABI method succeeded in accelerating the hologram calculation by approximately 30 times as compared to the conventional method.
We used a 3D object with a narrow depth of 3cm. 
In our upcoming project, we plan to perform a large-scale hologram calculation of 3D objects with deeper depths using the WASABI+WRP method \cite{wasabi2}.

\section*{Acknowledgments}
This work was partially supported by JSPS KAKENHI Grant Numbers 16K00151, and Ministry of Education, Culture, Sports, Science, and Technology (MEXT) Strategic Research Foundation at Private Universities (2013-2017).
We would like to thank Dr. Hirotaka Nakayama for providing 3D models.
This work was supported by Kan-Dai Digital Holo-Studio in Kansai University.

\bibliographystyle{model1a-num-names}
\bibliography{<your-bib-database>}







\end{document}